\documentclass[twocolumn,superscriptaddress,showpacs,aps,floatfix]{revtex4-1}
\usepackage{amsmath,amsfonts,amssymb,epsfig,dcolumn,bm,dsfont,graphics,latexsym,color,graphicx}
\usepackage{soul} 

\let\oldAA\AA

\def\10.1088/0022-3719/10/12/021ri{\mathrm{i}}
\renewcommand{\AA}{\text{\normalfont\oldAA}}
\newcommand{\ket}[1]{| {#1} \rangle} 
\newcommand{\bra}[1]{\langle {#1} |} 
\newcommand{\aver}[1]{\langle {#1} \rangle} 
\begin{document}
\preprint{AIP/123-QED}
\title{
Ferroelectric switching control of spin current in graphene proximitized by In$_2$Se$_3$}
\author{Marko Milivojevi\'c}
\email{marko.milivojevic@savba.sk}
\affiliation{Institute of Informatics, Slovak Academy of Sciences, 84507 Bratislava, Slovakia}
\affiliation {Faculty of Physics, University of Belgrade, 11001 Belgrade, Serbia}
\author{Juraj Mnich}
\email{juraj.mnich@student.upjs.sk}
\affiliation{Institute of Physics, Pavol Jozef \v{S}af\'{a}rik University in Ko\v{s}ice, 04001 Ko\v{s}ice, Slovakia}
\author{Paulina Jureczko}
\affiliation{Institute of Physics, University of Silesia in Katowice, 41‑500 Chorz\'ow, Poland}
\affiliation{Institute of Experimental Physics, Slovak Academy of Sciences, 04001 Ko\v{s}ice, Slovakia}
\author{Marcin Kurpas}
\affiliation{Institute of Physics, University of Silesia in Katowice, 41‑500 Chorz\'ow, Poland}
\author{Martin Gmitra}
\affiliation{Institute of Physics, Pavol Jozef \v{S}af\'{a}rik University in Ko\v{s}ice, 04001 Ko\v{s}ice, Slovakia}
\affiliation{Institute of Experimental Physics, Slovak Academy of Sciences, 04001 Ko\v{s}ice, Slovakia}
\affiliation{New Technologies Research Centre, University of West Bohemia, Univerzitní 8, CZ-301 00 Pilsen, Czech Republic}
\begin{abstract}
By utilizing the proximity effect, we introduce a platform that exploits ferroelectric switching to modulate spin currents in graphene proximitized by ferroelectric In$_2$Se$_3$ monolayer.
Through first-principles calculations and tight-binding modeling, we studied the electronic structure of graphene/In$_2$Se$_3$ heterostructure for twist angles of 0$^{\circ}$ and 17.5$^{\circ}$, considering both ferroelectric polarizations. 
We discover that switching the ferroelectric polarization reverses the sign of the charge-to-spin conversion coefficients, acting as a chirality switch of the in-plane spin texture in graphene. 
For the twisted heterostructure, we observed emergence of unconventional radial Rashba field for one ferroelectric polarization direction. 
Additionally, we demonstrated that the Rashba phase can be directly extracted from the ratio of conversion efficiency coefficients, providing a straightforward approach to characterize the in-plane spin texture in graphene. 
All the unique features of the studied graphene/In$_2$Se$_3$ heterostructure can be experimentally detected, offering a promising approach for developing advanced spintronic devices with enhanced performance and efficiency.
\end{abstract}
\maketitle
\section{Introduction}
\label{introduction}
Ferroelectricity in two-dimensional (2D) materials has emerged as a significant area of research due to its potential applications in next-generation electronics, including spintronics~\cite{ZFS04,FME+07}. This phenomenon refers to the ability of certain materials to exhibit spontaneous polarization that can be reversed by the application of an external electric field.

A recent computational study~\cite{KPG+23} has discovered a variety of 2D ferroelectric materials, including 49 with in-plane polarization and 8 exhibiting out-of-plane polarization. However, the list of experimentally synthetised 2D ferroelectric materials is not as long and consists of In$_2$Se$_3$~\cite{ZWZ+17,ZYZ+18,CHY+18,XZW+18}, CuInP$_2$Se$_6$~\cite{LYS+16,BNT+20},
MoTe$_2$~\cite{YLC+19}, NiI$_2$~\cite{SOE+22}, SnTe~\cite{CLL+16}, SnSe~\cite{CKM+20}, and SnS~\cite{HKL+20}. Those materials, besides hosting the ferroelectric phase, have a nonzero spin-orbit coupling (SOC), present due to the broken inversion symmetry.
The coexistence of ferroelectric and a SOC mediates interesting effects, such as the switching of spin-texture chirality with an electric-polarization reversal~\cite{KIF+14,MBS+16,KMB+16,RVA+18,WGG+20,GZW+23}.
A great application potential of the graphene/In$_2$Se$_3$ heterostructures has been demonstrated in neuromorphic devices with optical and electrical stimuli~\cite{MDG+23}. 

\begin{figure*}[t]
    \centering
    \includegraphics[width=0.825\linewidth]{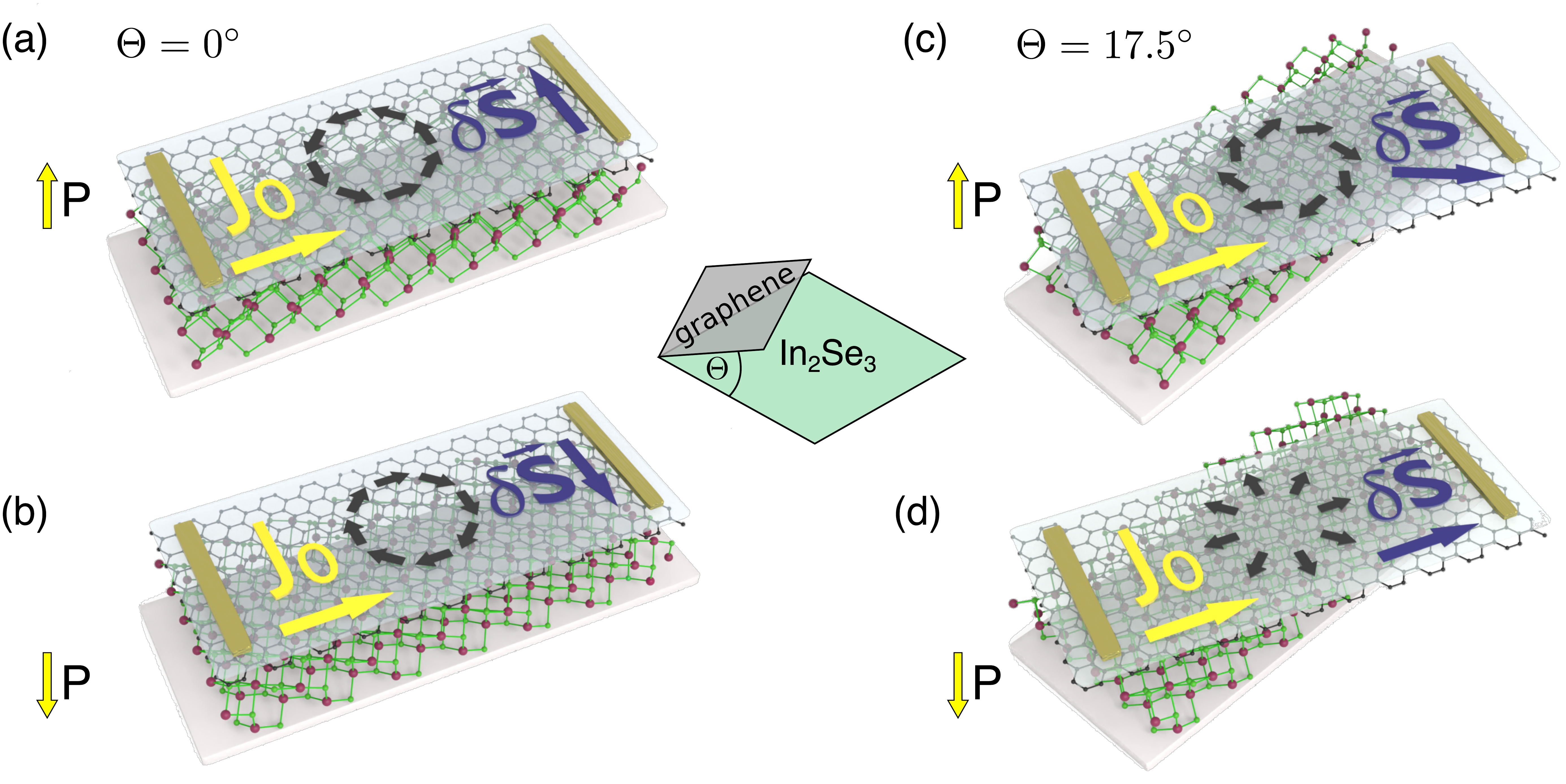}
    \caption{
    Operational features of the transport device made of graphene/In$_2$Se$_3$ heterostructure demonstrating chirality switch and unconventional Rashba Edelstein regimes.
    (a)~Sketch of the zero twist angle ($\Theta=0^\circ$) for positive polarization P$\uparrow$. The proximity-induced spin texture in graphene (black arrows) in spin-split bands near the $K$-point. The charge current $J_0$ (yellow arrow) of the hole-doped graphene flowing between the contacts (gold stripes) generates the current-induced nonequilibrium spin density $\delta\vec{S}$ (blue arrow) in the direction perpendicular to $J_0$ governed by the conventional Rashba-Edelstein effect.
    (b)~Upon flipping the polarization direction of In$_2$Se$_3$ from P$\uparrow$ ($+z$ direction) to P$\downarrow$ ($-z$ direction) the $\delta\vec{S}$ direction is switched. 
    (c)~For twist angle $\Theta=17.5^\circ$ and P$\uparrow$, the Rashba spin texture acquires a radial component resulting in a noncollinear $\delta\vec{S}$ orientation concerning the $J_0$ direction.
    (d)~For P$\downarrow$, $\delta\vec{S}$ is almost parallel to $J_0$, realizing the unconventional Rashba Edelstein effect.
    }
    \label{fig:device}
\end{figure*}

In this study, we demonstrate that the out-of-plane electric polarization of In$_2$Se$_3$ monolayer, having the electrically addressable vertical polarization~\cite{DZW+17,BKL+24}, can be used to control the spin current in graphene within the graphene/In$_2$Se$_3$ heterostructure.
We note that the spontaneous electric polarizations in In$_2$Se$_3$ monolayer are separated by the activation barrier of about 0.06~meV~\cite{DZW+17}.
Utilizing first-principles calculations, we investigated the graphene/In$_2$Se$_3$ heterostructure at twist angles of 0$^{\circ}$ and 17.5$^{\circ}$, accounting for both ferroelectric states.
We show that these two distinct configurations, characterized by different symmetries --  ${\bf C}_{3{\rm v}}$ in the zero-twist angle case and ${\bf C}_{3}$ in the 17.5$^{\circ}$ case -- result in different spin current responses in graphene. 
In FIG.~\ref{fig:device}, we sketch the transport device with a ferroelectric control of the charge-to-spin conversion in four regimes. At the zero twist angle, the polarization direction in In$_2$Se$_3$ acts as a {\it chirality switch} of the in-plane spin texture components of graphene.
The chirality switch can be verified by the sign switching of the spin current that can be detected in standard charge-to-spin conversion experiments utilizing Rashba-Edelstein effect (REE).
For a finite twist angle $17.5^\circ$ and negative electric polarization in In$_2$Se$_3$, the spin texture in graphene has mostly in-plane components with almost full radial Rashba spin texture.
In this case, the nonzero twist and negative electric polarization turn the device into an unconventional Rashba-Edelstein effect (UREE) regime.

The radial Rashba field~\cite{FFZ+24} has attracted significant interest because of its various applications~\cite{ZKK+20,ZPT+23,ZKF24b,KRL25}. Besides the unconventional charge-to-spin conversion protocols, it can lead to the improved spin-orbit torque functionality~\cite{HWW+21} and influence correlated phases~\cite{LZM+22,XS23,ZKF24} and superconductivity~\cite{ZPT+23,JSC+23}.
Thus, we believe that experimentally accessible graphene/In$_2$Se$_3$ heterostructure can bring new functionalities to the graphene-based van der Waals heterostructures~\cite{ATT+14,HKG+14,GF15,GF17,CGF+17,BST+18,ZCR+21,NZG+21,LSK+22,PDR+22,VPP+22,SMK+23,MGK+24,GAK+17,YLB+17,IHS+19,TF23,GVC+18,ZGF19,ZJN+23}.
Benefiting of graphene's long spin-relaxation times~\cite{SKX+16,DFP+16} and high electronic mobility~\cite{BSJ+08,DSB+08}, the graphene/In$_2$Se$_3$ heterostructure has a significant potential for spintronics applications.

The paper is organized as follows. In Sec.~\ref{first-principles} we present the first-principles calculation details of the heterostructures. The effective model Hamiltonian of proximitized graphene is given in Sec.~\ref{modelHam}, and the numerical procedure for the REE and UREE charge-to-spin conversion coefficients calculations is given in Sec.~\ref{charge-to-spin}. The main results are represented in Sec.~\ref{results}, and implications of the obtained results are discussed in Sec.~\ref{Conclusions}. 

\section{First-principles calculations}
\label{first-principles}
The sketch of a transport device made of graphene/In$_2$Se$_3$ heterostructure is shown in FIG.~\ref{fig:device}. 
We consider two terminal geometry with contacts for charge transport $J_0$ in graphene.
The capping layer, contacts, and substrate are included solely to visualize the device conceptually and are not considered in the first-principles calculations. 
The ferroelectric In$_2$Se$_3$ monolayer 
has two equivalent configurations, with the polarization $\mathbf{P}$
pointing in the positive P$\uparrow$ or negative P$\downarrow$ $z$-direction, see FIG.~\ref{fig:device}  and FIG.~\ref{fig:unfold}. 
The polarization direction plays a significant role when graphene is placed on In$_2$Se$_3$. The twist angle affects the electronic structure of graphene as well as its polarization.
We investigate two specific twist angles, $\Theta$=0$^{\circ}$ and $\Theta$=17.5$^{\circ}$ between the graphene and In$_2$Se$_3$ monolayers. In the zero twist angle case, the symmetry of the heterostructure is ${\bf C}_{3{\rm v}}$, whereas in the nonzero twist angle case, the overall symmetry is ${\bf C}_{3{\rm}}$.  
The lattice parameter of graphene is considered $a_0=2.46~{\rm\AA}$~\cite{HLS65}, and $4.106~{\rm\AA}$ for In$_2$Se$_3$~\cite{DZW+17}. 
The commensurate heterostructure for the twist angles is obtained by compressive deformation of the lattice constant of In$_2$Se$_3$. Using the latice vectors of graphene ${\bm a}_1=a_0{\bm e}_x$ and ${\bm a}_2=a_0(\cos{(2\pi/3)}{\bm e}_x+\sin{(2\pi/3)}{\bm e}_y)$ we can define lattice vectors of the heterostructure $(n_1,n_2)$ and $(m_1,m_2)$ as $(n_1,n_2)=n_1{\bm a}_1+n_2{\bm a}_2$ and $(m_1,m_2)=m_1{\bm a}_1+m_2{\bm a}_2$. In Table~\ref{TAB:structural} we list the considered lattice vectors $(n_1,n_2)$ and $(m_1,m_2)$, compressive strain $\delta$ to the In$_2$Se$_3$ lattice parameter, and the total number of atoms $N$ in the heterostructure for the twist angles $\Theta = 0^\circ$ and $17.5^\circ$. 

We perform the electronic structure calculation of the graphene/In$_2$Se$_3$ heterostructures using Density Functional Theory (DFT) as implemented in the plane wave code Q{\sc{uantum}} ESPRESSO (QE)~\cite{QE1,QE2}. For the exchange-correlation functional, we use the Perdew-Burke-Ernzerhof functional~\cite{PBE} within the projector augmented-wave method~\cite{pslib,PAW}. The kinetic energy cut-offs for the wave function and charge density were chosen to be 55~Ry and 326~Ry, respectively.
Methfessel–Paxton energy level smearing~\cite{MP89} of 1~mRy, and $12\times 12$ $k$-points mesh for the irreducible part of the Brillouin zone sampling were used for self-consistent calculations. 
The van der Waals interaction was modeled using the semiempirical Grimme-D2 correction~\cite{G06,BCF+08}, and a vacuum of 15~${\AA}$ in the $z$-direction to detach the periodic images of the heterostructure was used.
The positions of atoms were relaxed using the quasi-Newton scheme using scalar-relativistic pseudopotentials, keeping the force and energy convergence thresholds for ionic minimization to $1\times10^{-4}$~Ry/bohr and $10^{-7}$~Ry, respectively. 
The fully relativistic pseudopotentials were used for the self-consistent calculation, including the spin-orbit coupling, keeping the same $k$-mesh but increasing the energy convergence thresholds to $10^{-8}$~Ry. Also, dipole correction~\cite{B99} was applied to properly determine the Dirac point energy offset due to dipole electric field effects between graphene and In$_2$Se$_3$.
\begin{table}[t]
\caption{The structural information of the studied graphene/In$_2$Se$_3$ heterostructure. The table lists lattice vectors of the heterostructure $(n_1,n_2)=n_1{\bm a}_1+n_2{\bm a}_2$ and $(m_1,m_2)=m_1{\bm a}_1+m_2{\bm a}_2$ given in terms of the lattice vectors of the graphene lattice for two twist angles $\Theta$ between graphene and In$_2$Se$_3$,
applied compressive strain $\delta$ to the In$_2$Se$_3$ lattice parameter, and the total number of atoms $N$ in the heterostructure.}\label{TAB:structural}
\centering
\small
\setlength{\tabcolsep}{7pt}
\renewcommand{\arraystretch}{1.0}
\begin{tabular}{ccccc}
\hline\hline
$\Theta$ (deg) & $(n_1,n_2)$ & $(m_1,m_2)$ &  $\delta$ [\%] & $N$ \\\hline
 0 & $(5,0)$ & $(0,5)$ & $-0.15$ & 95 \\\hline
17.5$^\circ$ & $(2,5)$ &$(3,-2)$ & $-1.3$ & 73 \\\hline\hline
\end{tabular}
\end{table}

\section{Effective model Hamiltonian}
\label{modelHam}
The effective tight-binding Hamiltonian $H$ describing the low energy dispersion of graphene~\cite{KIF17} proximitized by In$_2$Se$_3$, having the ${\bf C}_{3{\rm v}}$ (${\bf C}_{3})$ symmetry for the zero (nonzero) twist angle, can be written as  $H=$ $H_0$+$H_{\rm I}$+$H_{\rm PIA}$+$H_{\rm{R}}(\phi_{\rm R})$. Here $H_0$ represents the orbital Hamiltonian, described in terms of the sublattice-dependent on-site 
potential, $\sum_{i=\rm{A},\rm{B}}\sum_{\sigma}(\mu+\Delta (-1)^{\delta_{\rm{B},i}})c_{i\sigma}^{\dagger}c_{i\sigma}$,
equal to $\mu\pm \Delta$ on sublattice A/B, where $\mu$ is the chemical potential, $\Delta$ is the stagerred potential, and $c_{i\sigma}^{\dagger}$ ($c_{i\sigma}$) is the creation (anihilation) operator on site $i=\rm{A},\rm{B}$ with spin $\sigma=\pm$. 
Furthermore, the orbital Hamiltonian consists of nearest-neighbor Hamiltonian with hopping $t$, $-t \sum_{\aver{m,n}}\sum_{\sigma}c_{n\sigma}^{\dagger}c_{m\sigma}$, which can be connected to the Fermi velocity $v_{\rm F}$ through the relation $v_{\rm F}=a t \sqrt{3}/2\hbar$. 
The second term $H_{\rm I}=\sum_{\alpha={\rm A,B}}\sum_{\aver{\aver{m,n}}}\frac{{\rm i}\lambda_{\rm I}^{\alpha}}{3\sqrt{3}}\sum_{\sigma}\nu_{m,n}[s_z]_{\sigma\sigma}c_{m\sigma}^{\dagger}c_{n \sigma}$ describes the intrinsic SOC, where the sum goes over the next-nearest-neighbors interaction described in terms of the sublattice-dependent $\lambda_{\rm I}^{\rm{A}/\rm{B}}$ parameters and the sign factor $\nu_{m,n}$ that has the value 1 ($-1$) when the next-nearest-neighbor hopping from site $m$ to site $n$ via the common nearest-neighbor encloses a clockwise (counterclockwise) path. The third term $H_{\rm PIA}$ describes the pseudospin inversion asymmetry (PIA) SOC term~\cite{GKF13} $H_{\rm PIA}=\sum_{\alpha=\rm{A},\rm{B}}\frac{2\lambda_{\rm PIA}^{\alpha}}{3}\sum_{\aver{\aver{m,n}}}\sum_{\sigma\neq\sigma'}
[i {\bf s}\times {\bf d}_{m,n}]_{\sigma\sigma'}c_{m\sigma}^{\dagger}c_{n \sigma'}$, describing the pseudospin-dependent next-nearest-neighbor hopping, where $\lambda_{\rm PIA}^{\rm{A}/\rm{B}}$ is the hopping strength. The last term, $H_{R}(\phi_{\rm R})$, describes the nearest-neighbor Rashba SOC term, which additionally depends on the Rashba angle $\phi_{\rm R}$ \cite{DRK+19,PDR+22} as follows $H_{\rm{R}}(\phi_{\rm R})=\frac{2{\rm i}\lambda_{\rm R}}{3}\sum_{\aver{m,n}}\sum_{\sigma\neq\sigma'}
U^{\dagger}(\phi_{\rm R})[{\bf s}\times {\bf d}_{m,n}]_{\sigma\sigma'}^z U(\phi_{\rm R})c_{m\sigma}^{\dagger}c_{n \sigma'}$, where ${\bf s}$ is the vector of Pauli matrices, $\lambda_{\rm R}$ represents the Rashba SOC strength, and ${\bf d}_{m,n}$ is the unit vector in the horizontal plane pointing from lattice site $n$ to the nearest-neighbor site $m$.
The unitary operator $U(\phi_{\rm R})=e^{-i s_z \phi_{\mathrm R}/2}$ appears due to non-zero twist angle.

\section{Charge-to-spin conversion efficiencies}
\label{charge-to-spin}
Charge-to-spin conversion represents an experimentally verifiable signature of the proximity-induced SOC~\cite{GCR17,GKB+19,HSI+20,KHA+20,HKZ+21,IGH+22,CSC+22,CLY+24}.
The presence of the Rashba SOC, i.e., the in-plane spin texture, is responsible for the appearance of both the REE~\cite{E90,DBD14,OMR+17,GAK+17} and the UREE~\cite{PDR+22,LSK+22,YMK+23,OSH+23}.
Whereas in the REE the accumulated spins are perpendicular to the charge current, in the UREE the spin density polarization is collinear to the applied electric current. Assuming that the charge current, $\delta J_x$, is applied in the $x$-direction 
(direction of the applied bias voltage $V_{\rm{b}}$), the REE and UREE efficiencies, $\alpha_{\rm REE}$, and $\alpha_{\rm UREE}$, can be defined as
\begin{eqnarray}
    \alpha_{\rm REE}&=&\frac{e v_{\rm{F}}}{\hbar}\frac{\delta S_y}{\delta J_x},     \\
    \alpha_{\rm UREE}&=&\frac{e v_{\rm{F}}}{\hbar}\frac{\delta S_x}{\delta J_x},  
\end{eqnarray}
where $\delta S_x$ ($\delta S_y$) is the current-induced nonequilibrium spin density along the $x$ ($y$) axis.

The coefficients $ \alpha_{\rm REE}$ and
$\alpha_{\rm UREE}$ were calculated using the Kubo formula~\cite{K56,K57} in the Smrčka-Středa formulation~\cite{SS77,CB01,BM20}. Assuming the weak disorder scattering described by the phenomenological parameter $\gamma$~\cite{LSK+22,BM20,FBM14,ZZF+17}, we defined the change $\delta \mathcal{O}$ of the physical quantity represented by the operator $\mathcal{O}$ as 
\begin{equation}
    \delta  \mathcal{O}=\frac{E}{4\pi^2}\int d^2{\bf k}[\chi_{\mathcal{O}}^{\rm surf}({\bf k})+
     \chi_{\mathcal{O}}^{\rm sea}({\bf k})],
\end{equation}
where $\mathcal{O}=\{S_{x/y},J_{x}\}$, $S_{x/y}=\frac{\hbar}2 s_{x/y}$, and $J_{x} = -ev_{x}$. 
The Fermi surface and Fermi sea susceptibilities, $\chi_{\mathcal{O}}^{\rm surf}$ and $\chi_{\mathcal{O}}^{\rm sea}$, are equal to~\cite{BM20}
\begin{eqnarray}
    \chi_{\mathcal{O}}^{\rm surf}({\bf k})&=&\frac{\hbar}{\pi}\gamma^2\times \label{eq:response_surface}\\
    &\times&\sum_{n,m} \frac{{\rm Re}\{
    \bra{n{\bf k}}\mathcal{O}\ket{m{\bf k}} 
    \bra{m{\bf k}}J_x\ket{n{\bf k}}\}}
    {[(\epsilon_{n{\bf k}}-E_F)^2+\gamma^2][(\epsilon_{m{\bf k}}-E_F)^2+\gamma^2]},\nonumber\\
      \chi_{\mathcal{O}}^{\rm sea}({\bf k})&=&\hbar
    \sum_{n,m \neq n} (f_{n,{\bf k}}-f_{m,{\bf k}})\times\nonumber\\
    &&\times\frac{{\rm Im}\{
    \bra{n{\bf k}}\mathcal{O}\ket{m{\bf k}} 
    \bra{m{\bf k}}J_x\ket{n{\bf k}}\}}
    {(\epsilon_{n{\bf k}}-\epsilon_{m{\bf k}})^2}\,,\label{eq:response_sea}
\end{eqnarray}
where $\epsilon_{n{\bf k}}$ and $\ket{n {\bf k}}$, $n=1,\dots,4$, represent eigenvalues and eigenvectors of the graphene Hamiltonian in the vicinity of the $K/K'$ points, and $f_{n,{\bf k}}$ is the Fermi function.
In all calculations of charge-to-spin interconversion coefficients presented in Section~\ref{Charge-To-Spin}, we performed the integration of the response functions~\eqref{eq:response_surface} and~\eqref{eq:response_sea} around the $K$ and $K'$ points.
This integration was carried out on a sufficiently large grid having a linear size of at least 1.4~$k_{\rm{max}}$, with the step $\Delta k\approx 5 \times 10^{-7}$~\AA$^{-1}$,
where the parameter $k_{\rm{max}}$ represents the maximum diameter of the Dirac cone of graphene for a chemical potential considered. 
All calculations were performed at an electronic temperature $k_{\rm B}T=0.01$~meV.
For the quasiparticle lifetime, defined as $\gamma=\hbar/2\tau$, we assume the relaxation time $\tau = 10^{-10}$~s~\cite{JLS+21}.
\begin{figure}
    \centering
\includegraphics[width=0.99\linewidth]{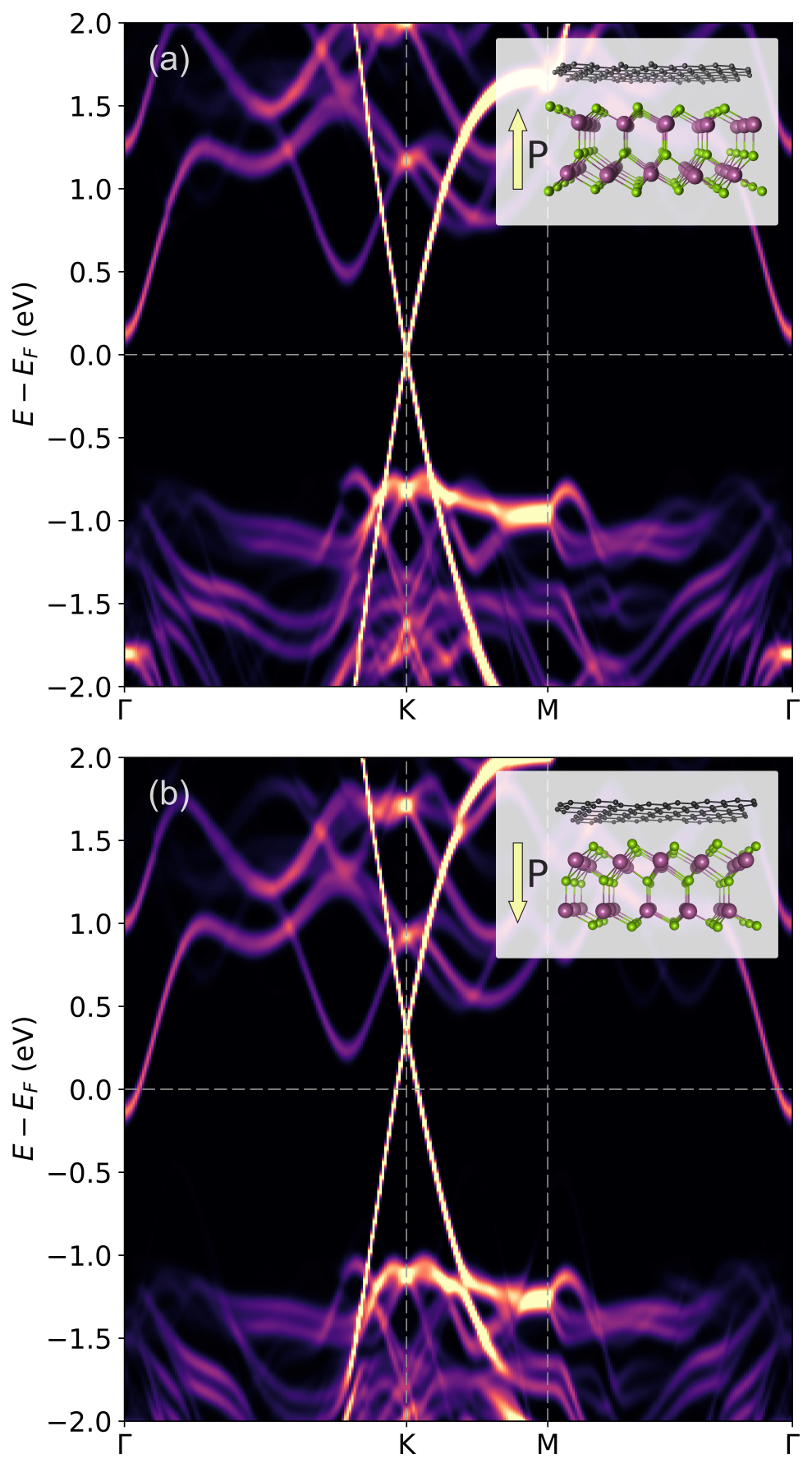}
    \caption{Calculated electronic structure of the twisted graphene/In$_2$Se$_3$ heterostructure ($\Theta=17.5^{\circ}$) unfolded to the first Brillouin zone of the graphene primitive cell. (a) Electronic states for the positive polarization P$\uparrow$ of In$_2$Se$_3$ pointing in the $+z$ direction; (b) for the polarization P$\downarrow$ in the $-z$ direction. 
    }\label{fig:unfold}
\end{figure}

\section{Results}
\label{results}
At the interface of two materials joined in a van der Waals heterostructure, the generated displacement field allows for highly efficient electric control of the proximity effects~\cite{GF17}.
The field modifies the electronic structure of the materials via doping and proximity-induced effects.
In the graphene/In$_2$Se$_3$ heterostructure, the displacement is compensated for the positive direction of the In$_2$Se$_3$ electric polarization P$\uparrow$, leading to a small chemical potential $\mu$ with electron doping of graphene. 
Negative polarization P$\downarrow$ adds to the displacement field, and significantly enhances the hole doping of graphene, see FIG.~\ref{fig:unfold}.

Structure inversion asymmetry of the graphene/In$_2$Se$_3$ heterostructure breaks the graphene lattice's horizontal mirror plane symmetry and removes its bands' spin degeneracy.
The spin splitting is governed by the SOC proximity-induced effects, and can be controlled by the direction of the electric polarization in In$_2$Se$_3$ layer.
To describe the proximity-induced effects in graphene as a function of the twist angle and polarization direction, we extract the effective tight-binding model parameters, described in Sec.~\ref{modelHam}, that fit the DFT data. The extracted parameters are given in Table~\ref{TAB:parameters}.
\begin{table}[htp]
\caption{Parameters of the effective tight-binding model of graphene on In$_2$Se$_3$ monolayer for the twist angles $\Theta=0^\circ$ and $\Theta=17.5^\circ$.}\label{TAB:parameters}
\centering
\small
\setlength{\tabcolsep}{7pt}
\renewcommand{\arraystretch}{1.0}
\begin{tabular}{rrr}
\hline\hline
 & Gr/In$_2$Se$_3$(P$\uparrow$)& Gr/In$_2$Se$_3$(P$\downarrow$) \\ \hline
& $\Theta=0^{\circ}$&\\\hline\hline
$v_{\rm F}$ [$10^{6}${\rm m/s}]&0.826 &0.822\\
$\mu$\;[{\rm meV}]&-0.921 &428.346\\
$\Delta$\;[{\rm meV}] &-0.181 &-0.255\\
$\lambda_{\rm I}^{\rm A}$\;[{\rm meV}] &0.098 &0.161\\
$\lambda_{\rm I}^{\rm B}$\;[{\rm meV}] &-0.092 &-0.107\\
$\lambda_{\rm R}$\;[{\rm meV}] &0.041 &-0.134\\
$\lambda_{\rm PIA}^{\rm A}$\;[{\rm meV}] &-0.050 &-1.337\\
$\lambda_{\rm PIA}^{\rm B}$\;[{\rm meV}] &1.940&1.516\\\hline\hline
& $\Theta=17.5^{\circ}$ &\\\hline\hline
$v_{\rm F}$ [10$^{6}${\rm m/s}]&0.825 &0.822\\
$\mu$\;[{\rm meV}]&-1.946 &313.928\\
$\Delta$\;[{\rm meV}] &-0.033 &-0.030\\
$\lambda_{\rm I}^{\rm A}$\;[{\rm meV}] &-0.002 &-0.014\\
$\lambda_{\rm I}^{\rm B}$\;[{\rm meV}] &0.003 &-0.005\\
$\lambda_{\rm R}$\;[{\rm meV}] &-0.174 &-0.100\\
$\phi_{\rm R}$\;[{\rm meV}] & 19.504 & 87.037\\
$\lambda_{\rm PIA}^{\rm A}$\;[{\rm meV}] &0.413 &-0.609\\
$\lambda_{\rm PIA}^{\rm B}$\;[{\rm meV}] &-0.198&0.709\\\hline\hline
\end{tabular}
\end{table} 

\begin{figure}[t]
    \centering
\includegraphics[width=0.99\linewidth]{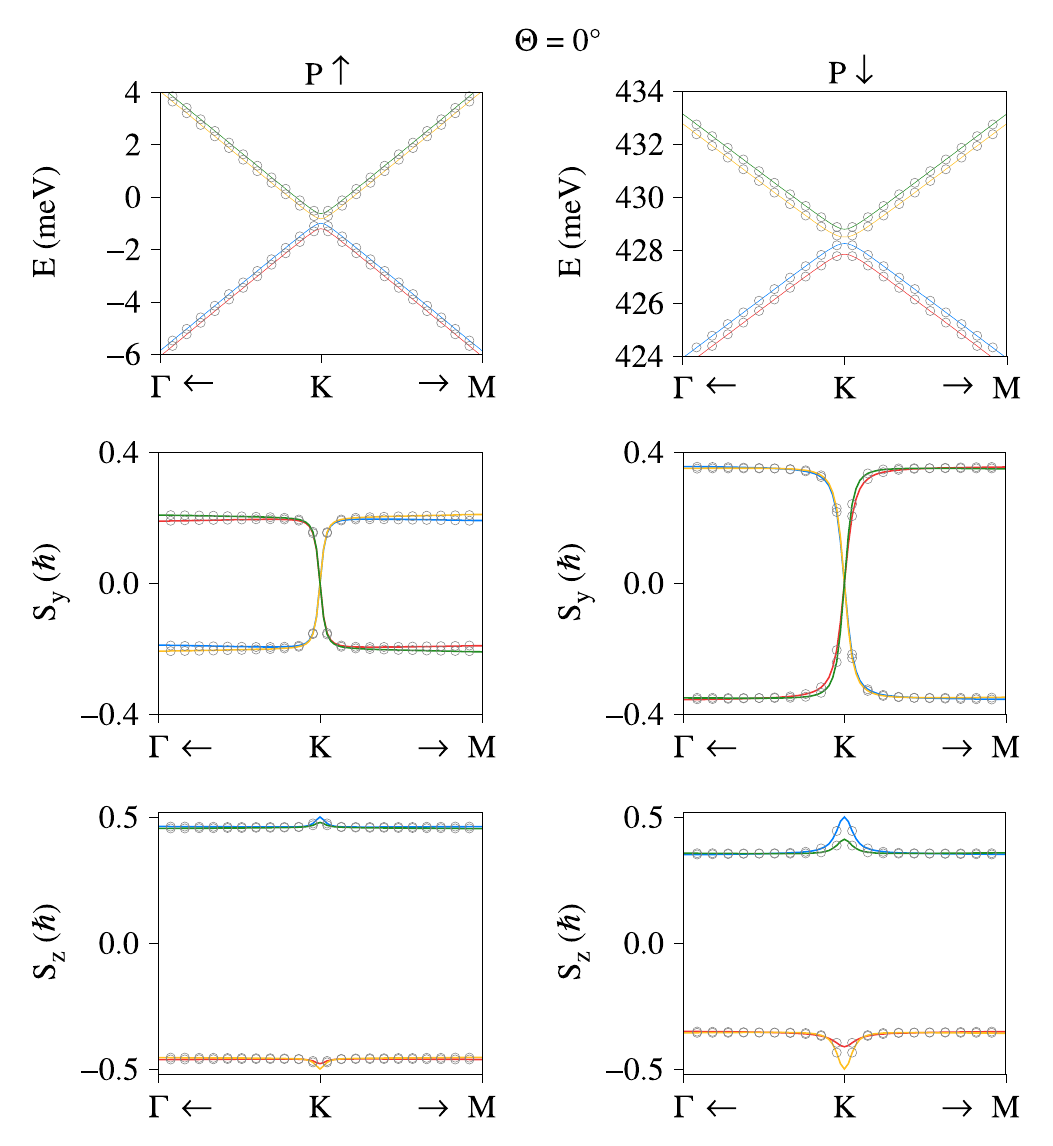}
    \caption{Calculated electronic band structure and spin expectation values close to the Dirac point of graphene on In$_2$Se$_3$ monolayer. The relative twist angle $\Theta=0^{\circ}$, polarization P$\uparrow$ (left panel) and P$\downarrow$ (right panel) of the In$_2$Se$_3$ monolayer were considered. The solid lines are a tight-binding model fit to the DFT data shown by circles. Each of the four line colors corresponds to a distinct band throughout the figure. We note that the $S_x$ expectation value is 0 by symmetry, because the $k$-path is taken along the $k_y=0$ line.} 
    \label{fig:fit0}
\end{figure}
\begin{figure}
    \centering
\includegraphics[width=0.99\linewidth]{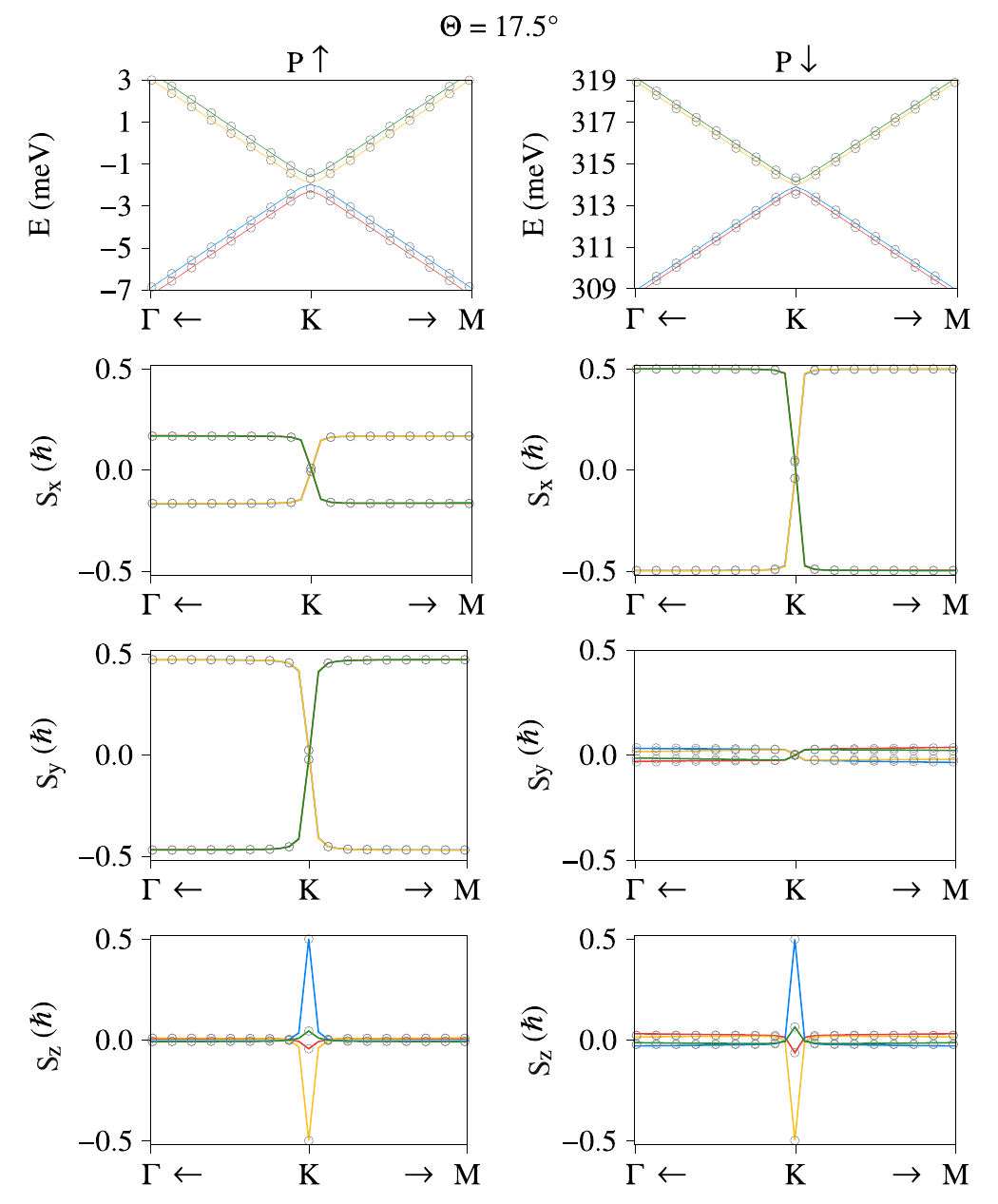}
    \caption{Calculated electronic band structure and spin expectation values close to the Dirac point of graphene on In$_2$Se$_3$ monolayer as described in FIG.~\ref{fig:fit0} but for the twist angle $\Theta=17.5^\circ$.} 
    \label{fig:fit175}
\end{figure}

The spin-independent parameters $v_{\rm F}$ and $\Delta$ are not affected much by the twist angle $\Theta$. However, there is a significant change in the chemical potential $\mu$, as expected from the large out-of-plane polarization of In$_2$Se$_3$, manifested as a Dirac cone shift relative to the Fermi level; see Table~\ref{TAB:parameters}.
The spin-dependent parameters can be split into two groups, those affecting the $S_z$ spin component, $\lambda_{\rm I}^{\rm{A}/\rm{B}}$, and those affecting the in-plane spin texture, such as $\lambda_{\rm R}$, $\phi_{\rm R}$ and $\lambda_{\rm PIA}^{\rm{A}/\rm{B}}$. 

In FIG.~\ref{fig:fit0} and \ref{fig:fit175} we compare the calculated electronic bands and spin expectation values of graphene on In$_2$Se$_3$ monolayer near the Dirac point from DFT and the tight-binding model.
The model accurately captures the qualitative trends observed in DFT calculations.

\begin{figure}[t]
    \centering
\includegraphics[width=0.99\columnwidth]{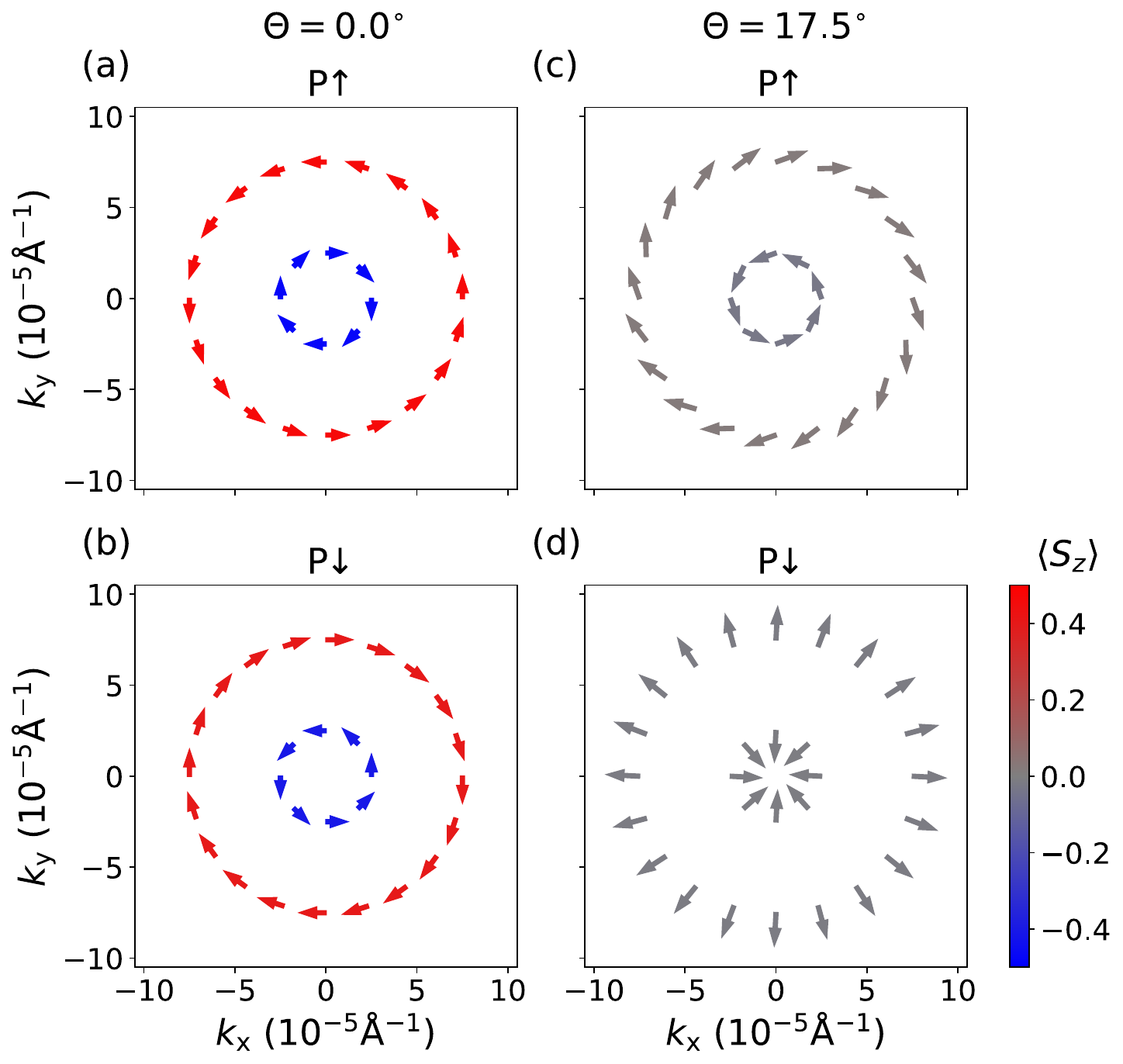}
    \caption{In-plane spin texture of the valence graphene bands close to the $K$ point calculated using the tight-binding model.
    (a)~Textures for positive polarization P$\uparrow$,
    (b)~for negative polarization P$\downarrow$ and zero twist angle.
    (c)~Spin textures for the twist angle $\Theta = 17.5^{\circ}$ and positive polarization P$\uparrow$,
    (d)~for negative polarization P$\downarrow$.
    The colors represent the spin $S_z$ expectation value. 
    The outer and inner circle radii of 7.5$\times 10^{-5}$~\AA$^{-1}$ and 2.5$\times 10^{-5}$~\AA$^{-1}$ correspond approximately
    to an energy of $-0.3$~meV relative to the chemical potential $\mu$. 
    } 
    \label{fig:spintexture}
\end{figure}

In FIG.~\ref{fig:spintexture}, the in-plane spin texture of graphene close to the $K$ point is depicted, for the studied cases.
We notice that around the $K$ point the dominant contribution to the overall SOC field is due to the Rashba and the intrinsic SOC terms, which are $k$-independent as opposed to the $k$-dependent PIA terms proportional to $\lambda_{\rm PIA}^{{\rm A/B}}$~\cite{KIF17}, that are usually neglected. Therefore, we focus on the contribution of the intrinsic and Rashba SOC only.

Starting with the zero twist angle case, we first notice that the polarization switching P$\uparrow \longrightarrow$P$\downarrow$ has only a quantitative change on the $\lambda_{\rm PIA}^{\rm{A}/\rm{B}}$ parameters, whereas it leads to the sign change of the Rashba spin-orbit field. 
As a consequence, there is a \textit{switch of the spin chirality} in graphene locked to the polarization direction. 
To explain this, we first notice that at the graphene/In$_2$Se$_3$ interface, the effective out-of-plane electric field in the $z$-direction, which drives the Rashba effect, arises from the interplay of two competing contributions: the proximity-induced electric field and the polarization-induced electric field. 
The proximity-induced field is an averaged quantity that depends on the distance between graphene and the nearest Se and In atomic planes, and therefore remains largely unaffected by the reversal of the In$_2$Se$_3$ polarization. In contrast, the polarization-induced field is directly tied to the direction of the ferroelectric polarization. Consequently, switching the polarization direction reverses the sign of the total electric field, which in turn switches the Rashba field. This behavior highlights the dominant role of the polarization-induced electric field in controlling the Rashba effect, consistent with expectations.

The non-zero twist angle between graphene and In$_2$Se$_3$ triggers the Rashba phase $\phi_{\rm R}$ and the possibility to realize unconventional spin-to-charge conversion protocols. 
By analyzing the effective parameters for the twist angle 17.5$^{\circ}$, we reveal that the out-of-plane intrinsic SOC is much weaker than the in-plane Rashba SOC contribution, because $|\lambda_{\rm I}^{\rm{A}/\rm{B}}|\ll |\lambda_{\rm R}|$. In addition to this, the polarization change P$\uparrow \longrightarrow$P$\downarrow$, enhances the Rashba phase drastically, introducing another {\it twistronics without a twist} control knob~\cite{SMK+23} that can be exploited in spintronics devices. 
A particularly interesting case, is the case of the heterostructure with negative polarization direction for which the Rashba phase $\phi_{\rm R}=87.037^{\circ}$ is obtained, suggesting an almost perfect radial Rashba field of $90^{\circ}$ that can enrich spin manipulation capabilities~\cite{HWW+21,LZM+22,XS23,ZKF24,ZPT+23,JSC+23}.

\subsection{Charge-to-spin conversion coeficients}\label{Charge-To-Spin}
\begin{figure}
    \centering
    \includegraphics[width=0.99\linewidth]{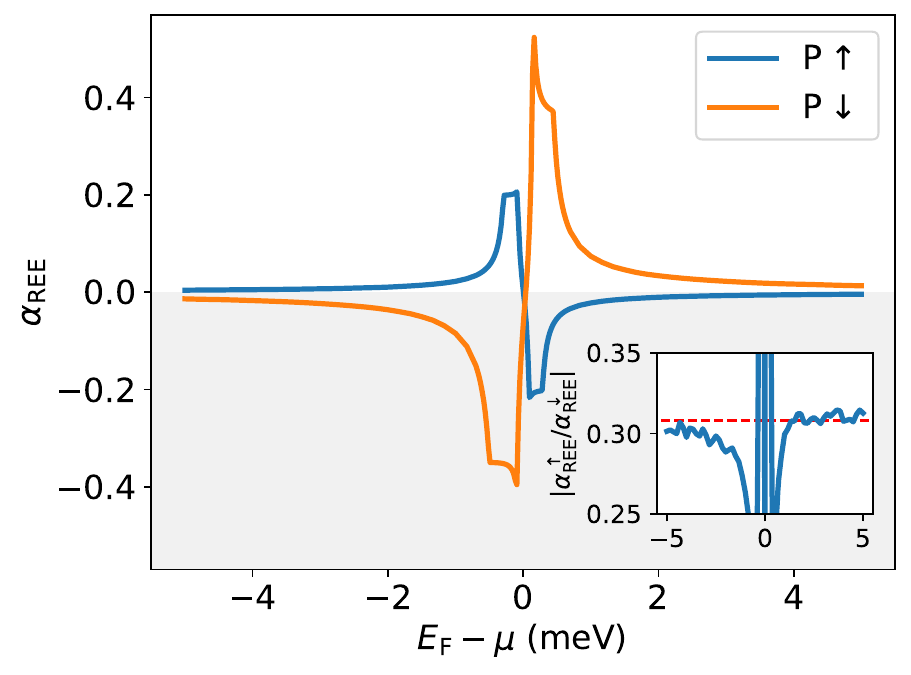}
    \caption{Calculated dependence of the conventional Rashba-Edelstein charge-to-spin conversion efficiency parameter $\alpha_{\rm{REE}}$ as a function of doping for both electric polarizations in In$_2$Se$_3$ monolayer for the twist angle $\Theta = 0^{\circ}$. The inset shows the ratio of considered coefficients for different polarizations $\alpha_{\rm{REE}}^{\uparrow}/\alpha_{\rm{REE}}^{\downarrow}$, where the red dashed line represents the estimated ratio from the parameters of the effective model.}
    \label{fig:(U)REE_0.0}
\end{figure}

To provide experimental implications of the proximity-induced effects in graphene, we study the efficiency of the charge-to-spin conversion coefficients.
FIG.~\ref{fig:(U)REE_0.0} shows $\alpha_{\rm REE}$ as a function of doping for $\Theta=0^{\circ}$ and two out-of-plane polarization directions in In$_2$Se$_3$.
In the zero twist angle case, only the Rashba-Edelstein effect is present due to ${\bf C}_{3{\rm v}}$ symmetry.
We note that the sign of $\alpha_{\rm REE}^{\uparrow}$ is opposite to the $\alpha_{\rm{REE}}^{\downarrow}$ for all doping level values, suggesting that the Rashba parameter change in the effective model can be directly connected to the sign change of $\alpha_{\rm REE}^{\uparrow\downarrow}$, i.e., switching of the spin current with the polarization direction reversal. Thus, this heterostructure holds substantial promise for nonvolatile control of the spin current direction in graphene. 

We mention that one can analyze the ratio of the conversion coefficients for different polarization states. We derived the following empirical relation 
$\alpha_{\rm{REE}}^{\uparrow}/\alpha_{\rm{REE}}^{\downarrow} \propto \lambda_{\rm{R}}^{\uparrow}/\lambda_{\rm{R}}^{\downarrow}$, depending solely on Rashba parameters.
The inset of FIG.~\ref{fig:(U)REE_0.0} shows the ratio $\alpha_{\rm{REE}}^{\uparrow}/\alpha_{\rm{REE}}^{\downarrow}$, where the red dashed line represents the calculated ratio using the empirical formula with Rashba parameters of the effective model.
The ratios of charge-to-spin interconversion coefficients for different polarization states are typically accompanied by a strong variation around the charge neutrality point. 
The main reason is different energy gaps for the P$\uparrow$ and P$\downarrow$ polarization states. The value from the empirical formula is subsequently reached at the higher doping levels.

\begin{figure}[t]
    \centering
\includegraphics[width=0.99\linewidth]{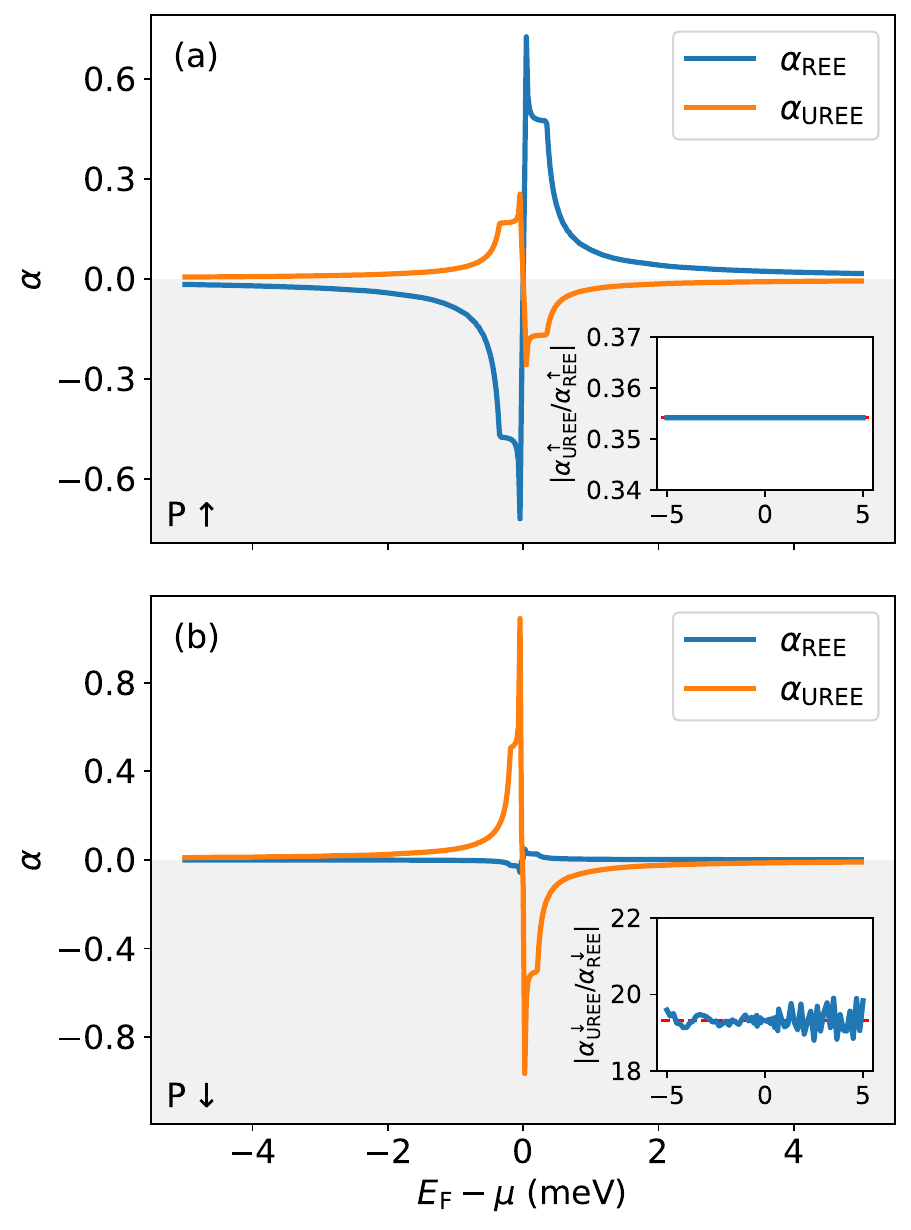}
   \caption{Conversion coefficients $\alpha_{\rm{REE}}$ and $\alpha_{\rm{UREE}}$ as a function of doping for the twist angle $\Theta = 17.5^{\circ}$ for 
   (a)~the positive electric polarization P$\uparrow$ and 
   (b)~the negative polarization P$\downarrow$ of the In$_{2}$Se$_{3}$ monolayer. 
   The insets show the ratio of unconventional and conventional Rashba-Edelstein efficiency parameters of the considered electric polarization state. The red dashed lines represent the estimated values of $\alpha_{\rm{UREE}}^{\uparrow\downarrow}/\alpha_{\rm{REE}}^{\uparrow\downarrow}$ ratio.}
    \label{fig:(U)REE_17.5}
\end{figure}

In FIG.~\ref{fig:(U)REE_17.5} we present the charge-to-spin conversion efficiency coefficients for a twist angle of $\Theta=17.5^{\circ}$.
In this case, the broken vertical mirror symmetry enables the generation of two spin current components: one orthogonal to the applied charge current, ($\propto \alpha_{\rm{REE}}$), and one parallel to it, proportional to ($\propto \alpha_{\rm{UREE}}$).
The inset of FIG.~\ref{fig:(U)REE_17.5} shows the
ratio $\alpha_{\rm{UREE}}^{\uparrow \downarrow}/\alpha_{\rm{REE}}^{\uparrow \downarrow}$ as a function of doping.
Our calculations reveal that this ratio is nearly independent of doping and can be directly related to the Rashba phase angle through the simple relation $\alpha^{\uparrow \downarrow}_{\rm{UREE}}/\alpha^{\uparrow \downarrow}_{\rm{REE}} \approx \tan (\phi_{\rm R}^{\uparrow \downarrow})$.
For positive polarization P$\uparrow$ we find  $\tan(\phi_{\rm{R}}^{\uparrow})\approx 0.354$,
indicating that the spin current is predominantly orthogonal to the charge current.
For negative polarization P$\downarrow$, $\tan(\phi_{\rm{R}}^{\downarrow})\approx 19.471$, which implies that the spin current aligns almost parallel to the charge current, consistent with a Rashba phase angle $\phi_{\rm R} = 87.037^\circ$.
These results confirm that the Rashba phase angle governs the direction of the spin current. Furthermore, they support the model assumption that ferroelectric switching can effectively control the spin current direction, achieving the radial Rashba limit in the  P$\downarrow$ polarization state.
Finally, our findings demonstrate that the Rashba phase can be directly extracted from the ratio of the unconventional and conventional Rashba–Edelstein effect coefficients, providing a straightforward experimental method to determine in-plane spin texture in graphene.

\section{Conclusions}
\label{Conclusions}
We demonstrated the potential of utilizing ferroelectric material In$_2$Se$_3$ to control spin currents in graphene, through the ferroelectric polarization state and proximity-induced effect. 
We also showed that graphene on In$_2$Se$_3$ exhibits distinct spin current modulation depending on the twist angle.
At a twist angle of $0^{\circ}$, we observed a conventional Rashba-Edelstein effect, where reversing the ferroelectric polarization 
results in a sign reversal of the charge-to-spin conversion coefficient. This indicates that the direction of the generated spin current in graphene is locked to the ferroelectric polarization of In$_2$Se$_3$ via the proximity-induced effects. 
Moreover, at the twist angle of 17.5$^{\circ}$, we showed that switching of the ferroelectric polarization significantly changes the Rashba phase, reaching the unconventional Rashba-Edelstein regime for negative electric polarization in In$_2$Se$_3$, with the Rashba phase of $\phi_{\rm R} \approx 87^{\circ}$, indicating an almost perfectly radial Rashba field.
Importantly, we show that for both ferroelectric polarization states in In$_2$Se$_3$, the Rashba phase can be quantitatively extracted from the ratio between the unconventional and conventional Rashba–Edelstein coefficients, providing a practical tool to experimentally characterize graphene’s in-plane spin texture.
These findings present a promising opportunity for developing advanced spintronics devices that leverage the proximity-induced manipulation of spin texture through ferroelectricity.

\acknowledgments
M.M. acknowledges the financial support provided by the Ministry of Science, Technological Development and Innovations of the Republic of Serbia (Grant No. 451-03-9/2021-14/200162). This project has received funding from the European Union’s Horizon 2020 Research and Innovation Programme under the Programme SASPRO 2 COFUND Marie Sklodowska-Curie Grant Agreement No. 945478 and has been funded by the EU NextGenerationEU through the Recovery and Resilience Plan for Slovakia under the Project No. 09I05-03-V02-00055.
J.M.~acknowledges the EU NextGenerationEU through the Recovery and Resilience Plan for Slovakia under the project No. 09I03-03-V05-00008.
P.J.~acknowledges financial support provided by the Slovak Academy of Sciences project IMPULZ IM-2021-42. 
M.K.~acknowledges financial support provided by the National Center for Research and Development (NCBR) under the V4-Japan project BGapEng V4-JAPAN/2/46/BGapEng/2022.
M.G.~acknowledges financial support provided by the Slovak Research and Development Agency under Contract No. APVV-SK-CZ-RD-21-0114 and by the Ministry of Education, Research, Development and Youth of the Slovak Republic, provided under Grant No. VEGA 1/0104/25 and the Slovak Academy of Sciences project IMPULZ IM-2021-42, and support of the QM4ST project funded by Programme Johannes Amos Commenius, call Excellent Research (Project No. CZ.02.01.01/00/22\_008/0004572).

\end{document}